\documentclass[iop,revtex4]{emulateapj}
\usepackage{microtype}
\usepackage{graphics,graphicx}
\usepackage{helvet}
\usepackage[utf8]{inputenc}
\usepackage[T1]{fontenc}
\usepackage{soul}
\usepackage{hyperref}
\usepackage{amsmath, amssymb, textcomp, gensymb}
\usepackage{multirow}
\usepackage{fancyhdr}
\usepackage{natbib}
\usepackage{xspace}
\usepackage{color}
\usepackage[left]{lineno}

\def\arcmin {^{\prime}\xspace}
\def\arcsec {^{\prime\prime}\xspace}

\def\kms {km\,s$^{-1}$\xspace}

\def\mjybm {mJy\,beam$^{-1}$\xspace}
\def\ujybm {$\mu$Jy\,beam$^{-1}$\xspace}
\def\mjybmvert {$\left(\frac{\textrm{mJy}}{\textrm{beam}}\right)$\xspace}

\def\etal {\textit{et al.}\xspace}

\def\apj{ApJ}
\def\mnras{MNRAS}
\def\nat{Nat}

\def\araa{ARA\&A}                
\def\aap{A\&A}                   
\def\apjs{ApJS}                  
\def\pasp{PASP}                  
\def\apjl{ApJ}                   
\def\jqsrt{JQSRT}                

\begin{document}


\slugcomment{\textit{Accepted for publication in ApJL on 19 April 2017}}

\title{Unveiling the Role of the Magnetic Field at the Smallest Scales of Star Formation}
\shorttitle{Unveiling the Role of the Magnetic Field at the Smallest Scales of Star Formation}

\author{Charles L. H. Hull\altaffilmark{1,10}} 
\author{Philip Mocz\altaffilmark{1}}
\author{Blakesley Burkhart\altaffilmark{1}}
\author{Alyssa A. Goodman\altaffilmark{1}}
\author{Josep M. Girart\altaffilmark{2}}
\author{Paulo C. Cort\'es\altaffilmark{3,4}}
\author{Lars Hernquist\altaffilmark{1}}
\author{Volker Springel\altaffilmark{5,6}}
\author{Zhi-Yun Li\altaffilmark{7}}
\author{Shih-Ping Lai\altaffilmark{8,9}}


\shortauthors{Hull \etal}
\email{chat.hull@cfa.harvard.edu}


 \altaffiltext{1}{Harvard-Smithsonian Center for Astrophysics, 60 Garden Street, Cambridge, Massachusetts 02138, USA}
 \altaffiltext{2}{Institut de Ci\`encies de l'Espai, (CSIC-IEEC), Campus UAB, Carrer de Can Magrans S/N, 08193 Cerdanyola del Vall\`es, Catalonia, Spain}
 \altaffiltext{3}{National Radio Astronomy Observatory, Charlottesville, VA 22903, USA}
 \altaffiltext{4}{Joint ALMA Office, Alonso de C\'ordova 3107, Vitacura, Santiago, Chile}
 \altaffiltext{5}{Heidelberger Institut f\"ur Theoretische Studien, Schloss-Wolfsbrunnenweg 35, 69118 Heidelberg, Germany}
 \altaffiltext{6}{Zentrum f\"ur Astronomie der Universit\"at Heidelberg, Astronomisches Recheninstitut, M\"onchhofstr. 12-14, 69120 Heidelberg, Germany}
 \altaffiltext{7}{Department of Astronomy, University of Virginia, Charlottesville, VA 22903, USA}
 \altaffiltext{8}{Institute of Astronomy and Department of Physics, National Tsing Hua University, 101 Section 2 Kuang Fu Road, 30013 Hsinchu,
Taiwan}
 \altaffiltext{9}{Academia Sinica Institute of Astronomy and Astrophysics, PO Box 23-141, 10617 Taipei, Taiwan}
 \altaffiltext{10}{Jansky Fellow of the National Radio Astronomy Observatory}


\clearpage

\begin{abstract}
We report Atacama Large Millimeter/submillimeter Array (ALMA) observations of polarized dust emission from the protostellar source Ser-emb 8 at a linear resolution of 140\,AU.  Assuming models of dust-grain alignment hold, the observed polarization pattern gives a projected view of the magnetic field structure in this source.  Contrary to expectations based on models of strongly magnetized star formation, the magnetic field in Ser-emb 8 does not exhibit an hourglass morphology. Combining the new ALMA data with previous observational studies, we can connect magnetic field structure from protostellar core ($\sim$\,80,000\,AU) to disk ($\sim$\,100\,AU) scales. We compare our observations with four magnetohydrodynamic gravo-turbulence simulations made with the AREPO code that have initial conditions ranging from super-Alfv\'enic (weakly magnetized) to sub-Alfv\'enic (strongly magnetized).  These simulations achieve the spatial dynamic range necessary to resolve the collapse of protostars from the parsec scale of star-forming clouds down to the $\sim$\,100\,AU scale probed by ALMA.  Only in the very strongly magnetized simulation do we see both the preservation of the field direction from cloud to disk scales and an hourglass-shaped field at <\,1000\,AU scales.  We conduct an analysis of the relative orientation of the magnetic field and the density structure in both the Ser-emb 8 ALMA observations and the synthetic observations of the four AREPO simulations. We conclude that the Ser-emb 8 data are most similar to the weakly magnetized simulations, which exhibit random alignment, in contrast to the strongly magnetized simulation, where the magnetic field plays a role in shaping the density structure in the source.  In the weak-field case, it is turbulence---not the magnetic field---that shapes the material that forms the protostar, highlighting the dominant role that turbulence can play across many orders of magnitude in spatial scale. \\
\end{abstract}


\section{Introduction}
\label{sec:intro}

The interstellar magnetic field is predicted to be a primary regulator of the formation of stars \citep{Mestel1956,Shu1987,McKee1993,McKee2007}.  In the ``strong-field'' mode of star formation, the field is relatively unaffected by rotation or turbulence, and the gravitational collapse of the star-forming material is regulated by a strong, ordered magnetic field that has been inherited from the interstellar field \citep{Fiedler1993}.  Furthermore, the collapse of strongly magnetized dense gas ($>10^4$\,cm$^{-3}$) is predicted to pinch the magnetic field into an hourglass shape that persists down to scales $<100$\,AU \citep{Fiedler1993, Allen2003}.  

Do observations support this theoretical picture of star formation?
Some observational studies \citep{HBLi2009} support the strong-field model, finding that the orientation of the large-scale interstellar magnetic field, as traced by the polarization produced by aligned dust grains, is preserved all the way down to the scale where individual stars form.  Regarding the predicted hourglass structure, high-resolution polarimetric observations of low-mass protostars have either shown evidence of hourglass magnetic field morphologies or lacked the sensitivity or resolution to rule them out on $\lesssim$\,10,000\,AU scales \citep{Girart2006, Matthews2009, Rao2009, Dotson2010, Stephens2013, Hull2013, Hull2014}.  However, the unprecedented sensitivity of the Atacama Large Millimeter/submillimeter Array (ALMA) now allows us to make well-resolved maps of the magnetic field toward more than just the brightest few sources in the sky, enabling us to determine whether or not a given source has formed in a strongly magnetized environment.

The observational understanding of magnetic fields in star forming regions is complicated by the fact that magnetic fields in the interstellar medium are very difficult to measure.
Interstellar dust grains are aligned with their short axes parallel to magnetic field lines \citep{Lazarian2007}.  Grains are aligned by ``radiative torques'' \citep{Hoang2009}; in the interstellar medium, this radiation source is the interstellar radiation field, whereas in deeply embedded sources, the radiation source is a central protostar.  This preferential alignment causes the thermal emission from the dust to become partially polarized in the direction perpendicular to the magnetic field; thus, a map of polarized emission can give a projected view of magnetic field structure.

The understanding that giant molecular clouds are magnetized and turbulent \citep{Larson1981, Burkhart2015} has inspired a number of simulations of star forming regions treating the first stage of the collapse process in the context of a turbulently driven, self-gravitating, isothermal, ideal magnetohydrodynamic (MHD) fluid  \citep{Kritsuk2007,Collins2012,Federrath2013,Burkhart2015,Li2015}. Such an approach incorporates the physical processes that are dominant during the initial parsec-scale phase of collapse; radiation, feedback, and non-ideal MHD terms (such as, e.g., turbulent magnetic reconnection: \citealt{SantosLima2012, Seifried2012b, Leao2013}) may become important at later stages.

Previously, statistical studies of the importance of the magnetic field in individual sources were limited to extremely bright, high-mass sources that have a statistically significant number of polarization detections across the source \citep[e.g., studies of the interplay between turbulence and the magnetic field in star-forming regions:][]{Hildebrand2009, Houde2009, Houde2011, Houde2016}.  However, advances in both observational capabilities and high-resolution, high-dynamic-range simulations have brought us to the point where we can make robust, meaningful comparisons of observations and simulations even toward faint low-mass objects, thereby achieving a deeper understanding of the relevance of the magnetic field in a turbulent star-forming environment.  

In this Letter, we report ALMA observations of polarized dust emission from the protostellar source Ser-emb 8 at a linear resolution of 140\,AU.   
Ser-emb~8 is one of several low-mass, Class 0 protostars in the Serpens Main star-forming region that were observed in full polarization with the Combined Array for Millimeter-wave Astronomy (CARMA) as part of the TADPOL survey \citep{Hull2014}.  Typical of a young, embedded protostar, Ser-emb~8 has a clear bipolar outflow visible in both CO\,($J = 2 \rightarrow 1$) and SiO\,($J = 5 \rightarrow 4$).  The source's name comes from \citet{Enoch2009b, Enoch2011}, who observed Ser-emb~8 (also known as S68N) with Bolocam, estimating that the source has a mass of approximately 9.4\,$M_\odot$ within a 100,000\,AU region surrounding the protostar.

We compare our observations with four MHD gravo-turbulence simulations run with the AREPO code using super-Alfv\'enic (weakly magnetized) and sub-Alfv\'enic (strongly magnetized) initial conditions. Our simulations allow us to resolve the collapse process down to the $\sim$\,100\,AU scale probed by ALMA (the smallest computational cell size is a few AU), and to investigate the effects of changing the relative importance of magnetic and turbulent energies. The simulations can be used to create mock observations that we can compare with real observations, and from which we can infer the physical parameters that describe the turbulent, collapsing medium.  

This Letter is organized as follows: in Section \ref{sec:obs} we describe our observing strategy for the ALMA observations of the protostellar source Ser-emb 8.   In Section \ref{sec:sims} we describe the setup of the four MHD gravo-turbulence simulations run with the AREPO code using weakly and strongly magnetized initial conditions.
In Section \ref{sec:results} we compare the ALMA observations and AREPO simulations both visually and statistically using the histogram of relative orientation (HRO) technique.
Finally, we discuss our results in Section \ref{sec:disc} and offer concluding remarks in Section \ref{sec:conc}.

\vspace{1em}
\section{ALMA Observations}
\label{sec:obs}
The 870\,\micron{} (Band 7) ALMA dust polarization observations of Ser-emb 8 were taken on 2015 June 3 and 7, and have $\sim$\,0.3$\arcsec$ angular resolution.  The ALMA polarization data comprise 8\,GHz of wide-band dust continuum ranging in frequency from $\sim$\,336--350\,GHz, with a mean frequency of 343.479\,GHz.   For further details on the ALMA polarization system and the reduction of ALMA polarization data, see \citet{Cortes2016, Nagai2016}, 
\sloppy
as well as the Common Astronomy Software Applications (\texttt{CASA}) Guide on 3C 286 polarization using ALMA 1\,mm Science Verification data.\footnote{ALMA CASA Guide on 3C 286 polarization at 1\,mm: \textbf{\href{https://casaguides.nrao.edu/index.php/3C286_Polarization}{\texttt{https://casaguides.nrao.edu/index.php/3C286\_Polarization}}}}

\begin{figure*}
\begin{center}
\includegraphics[width=0.9\textwidth, clip, trim=0cm 2cm 0cm 0cm]{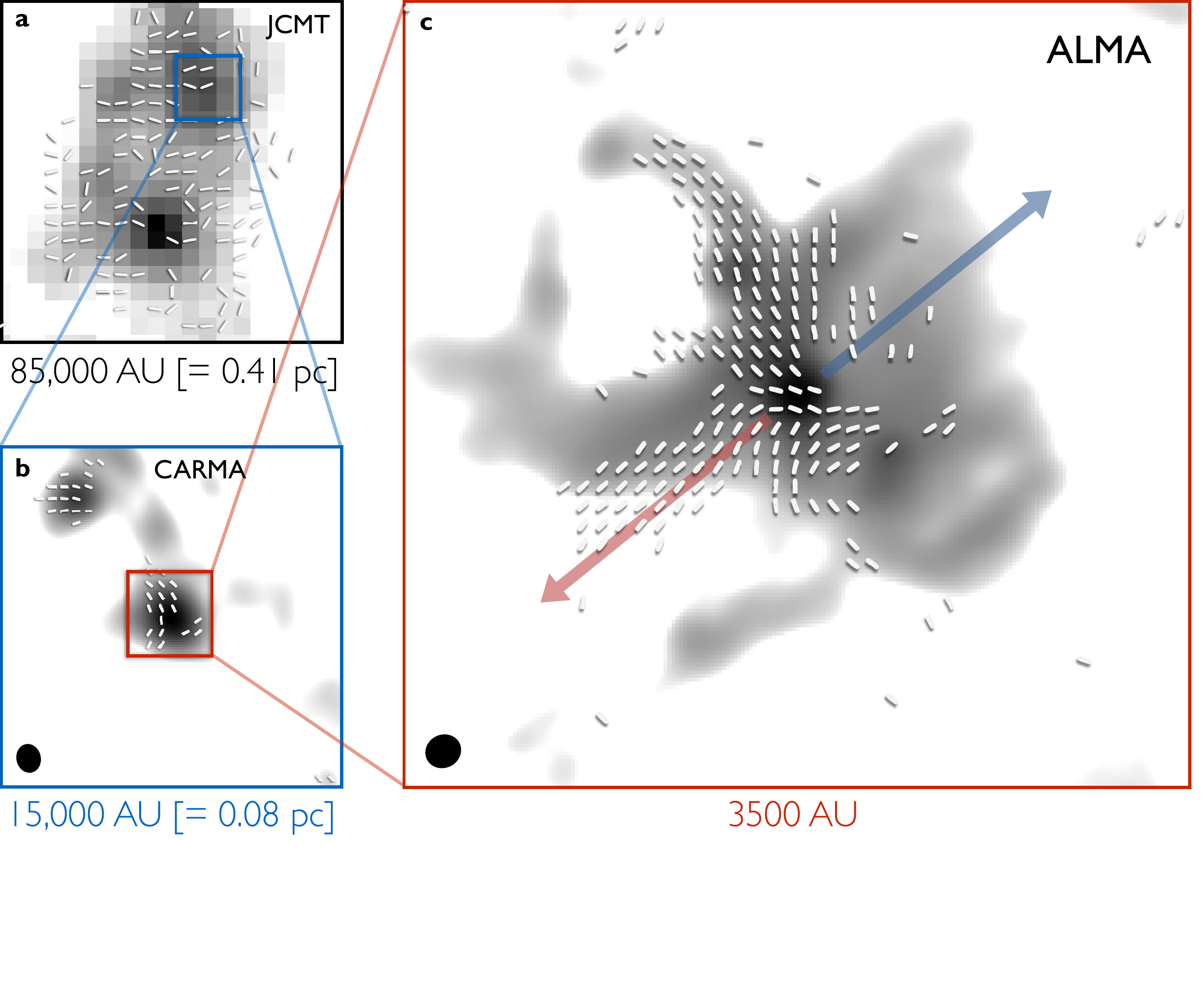}
\end{center}
\vspace{-3em}
\caption{\footnotesize Multi-scale view of the magnetic field around Ser-emb~8 ($\alpha_{\textrm{J2000}} =$ 18:29:48.089, $\delta_{\textrm{J2000}} =$ +1:16:43.32). 
Line segments represent the magnetic field orientation, rotated by 90$\degree$ from the dust polarization (the length of each segment is identical, and does not represent any other quantity).
Grayscale is total intensity (Stokes $I$) thermal dust emission.  
Panel (a) shows 870\,\micron{} JCMT observations \citep{Matthews2009}, (b) shows 1.3\,mm CARMA observations \citep{Hull2014}, and (c) shows 870\,\micron{} ALMA observations, revealing the magnetic field morphology with $\sim$\,10,000, 1000, and 140\,AU resolution, respectively. For the ALMA data, line segments are plotted where the polarized intensity $P > 3\sigma_P$; the rms noise in the polarized intensity map $\sigma_P = 25$\,\ujybm.  The dust emission is shown starting at 3 $\times$ $\sigma_I$, where the rms noise in the Stokes $I$ map $\sigma_I$ = 50\,\ujybm.  The peak polarized and total intensities in the ALMA data are 0.693\,\mjybm and 102\,\mjybm, respectively (the two peaks do not coincide exactly).  The red and blue arrows indicate the red- and blue-shifted lobes of the bipolar outflow \citep{Hull2014}. 
The text below each of the panels indicates the physical size of the image at the 436\,pc distance to the Serpens Main region (\citealt{OrtizLeon2017b}; see earlier results by \citealt{Dzib2010, Dzib2011}).  
The black ellipses in the lower-left corners of the ALMA and CARMA maps represent the synthesized beams (resolution elements).  The ALMA beam measures $0\farcs35\times0\farcs32$ at a position angle of --63$\degree$; the CARMA beam measures $2\farcs89\times2\farcs43$ at a position angle of 13$\degree$.  The JCMT data have a resolution of $20\arcsec$. 
\textit{The ALMA data used to make the figure in panel (c) are available in the online version of the ApJ Letter.}
\vspace{1em}
}
\label{fig:3panel}
\end{figure*}

\fussy
The dust continuum image (Figure \ref{fig:3panel}(c)) was produced by using the CASA task \texttt{CLEAN} with a Briggs weighting parameter of robust\,=\,1.  The final synthesized beam is $0\farcs35\times0\farcs32$ at a position angle of --61$\degree$.  The image was improved iteratively by four rounds of phase-only self calibration using the Stokes $I$ image as a model.  The Stokes $I$, $Q$, and $U$ maps were each \texttt{CLEANed} independently with an appropriate number of \texttt{CLEAN} iterations after the final round of self-calibration.  The rms noise level in the final Stokes $I$ dust map is $\sigma_I = 50$\,\ujybm{}, whereas the rms noise level in the Stokes $Q$ and $U$ dust maps is $\sigma_Q \approx \sigma_U \approx \sigma_P = 25$\,\ujybm{}; the reason for this difference is that the total intensity emission (Stokes $I$) image is more dynamic-range limited than the polarized emission (Stokes $Q$ and $U$).

The polarized intensity $P = \sqrt{Q^2 + U^2}$, the fractional polarization $P_\textrm{frac} = P / I$, and the polarization position angle $\chi = 0.5\,\arctan{\left(U / Q\right)}$.  Note that $P$ has a positive bias because the $P$ is always positive, even though the Stokes parameters $Q$ and $U$ from which $P$ is derived can be either positive or negative.  This bias has a particularly significant effect in low signal-to-noise measurements.  We thus debias the polarized intensity map as described in \citet{Vaillancourt2006} and \citet{Hull2015b}.  See Table \ref{table:data} for the ALMA polarization data.


\vspace{1em}
\section{AREPO simulations}
\label{sec:sims}
We perform our simulations using the AREPO code \citep{2010MNRAS.401..791S}, which uses a moving Voronoi mesh that follows the gas flow.  The mass in each gas cell remains approximately constant and automatically adapts to the geometry of the physical system.  The ability to accurately simulate the properties of the magnetic field, including maintaining the divergence-free condition, has only recently been developed in such codes \citep{Mocz2016}.

By running four turbulent, self-gravitating AREPO simulations of $\sim$\,5\,pc regions of star-forming clouds with different initial ratios of magnetic to turbulent energy, we investigate the role of the magnetic field in shaping the material immediately surrounding the protostar. The simulations are representative of typical star forming environments. The four simulations have different initial magnetic field strengths---and thus different Alfv\'en Mach numbers $\mathcal{M}_{\rm A} \equiv v_{\rm turb} / v_{\rm A}$---which are listed in Table \ref{table:sims}.  $v_{\rm turb}$ is the velocity of the driven turbulent motions in the box, and $v_{\rm A}\equiv B/\sqrt{4\pi\rho}$ is the Alfv\'enic wave speed.  

\begin{table*}
\normalsize
\begin{center}
\caption{\normalsize \vspace{0.1in} Initial parameters of the four simulations carried out with AREPO}
\begin{tabular}{cccccc}
\hline
\hline
sim. & $\beta_{\textrm{mean-field}}$ & $B_{\textrm{mean-field}}$ ($\micro$G) & $\mathcal{M}_{{\rm A},\textrm{mean-field}}$ & $\mathcal{M}_{{\rm s}}$ & Comment\\
\hline
1 & 25 & 1.2 & 35 & 10 & very weak field (super-Alfv\'enic)  \\
2 & 0.25 & 12 & 3.5 & 10 & weak field (super-Alfv\'enic) \\
3 & 0.028 & 36 & 1.2 & 10 & moderate field (trans-Alfv\'enic) \\
4 & 0.0025 & 120 & 0.35 & 10 & strong field (sub-Alfv\'enic) \\
\hline
\end{tabular}
\label{table:sims}
\end{center}
\footnotesize
\vspace{-0.5em}
\textbf{Note.} $\beta_{\textrm{mean-field}}$ indicates the initial plasma $\beta$, i.e., the ratio of gas pressure to magnetic pressure. 
$B_{\textrm{mean-field}}$ is the initial magnetic field strength in the 5.2\,pc box.
$\mathcal{M}_{{\rm A},{\textrm{mean-field}}}$ indicates the initial Alfv\'en Mach number, 
and $\mathcal{M}_{{\rm s}} \equiv v_{\textrm{turb}}/c_s$ is the initial sonic Mach number.
\vspace{1em}
\end{table*}

AREPO's base scheme solves the equations of ideal hydrodynamics with a finite-volume approach using a second-order unsplit Godunov scheme. In order to maintain the divergence-free property of the magnetic field on an unstructured mesh, we have implemented a constrained transport solver in terms of the magnetic vector potential to evolve the equations of ideal magnetohydrodynamics \citep{Mocz2016}. The method uses a Harten-Lax-van Leer-Discontinuities (HLLD) Riemann solver to accurately capture shocks. The moving-mesh method greatly reduces advection errors compared with traditional adaptive refinement mesh methods due to its quasi-Lagrangian nature. We also couple self-gravity to the MHD equations, which is calculated using a Tree-Particle-Mesh scheme. Solenoidal turbulence is driven in Fourier space at the largest spatial scales using an Ornstein-Uhlenbeck process \citep{Federrath2013}.

We simulate the collapse of star-forming cores in a turbulently driven interstellar medium. 
Our domain is a periodic box with side length $L_0 = 5.2~{\rm pc}$ and total mass $M=8000~{M_\odot}$.  
We use an isothermal equation of state, and the sound speed is set to $0.2$\,\kms.
We simulate initial field strengths $B_0=1.2,12,36,120\,\mu{\rm G}$, corresponding to plasma-betas ($\beta = P_{\rm gas}/P_{\rm B} = 8\pi\,P_{\rm gas}/B^2 $) of $\beta = 25,\,0.25,\,0.028,\,0.00025$. The initial magnetic field points in the vertical direction.  This mean-field value is an invariant of ideal MHD. The range of magnetic field strengths we consider includes fields that are both weak and strong relative to the gas pressure and turbulent pressure. Each simulation uses $256^3$ particles, which corresponds to a mass resolution of $8 \times 10^{-5}~{M_\odot}$ with our quasi-Lagrangian code. The simulations resolve collapse down to the scales of a few AU, giving an effective resolution in these collapsed regions of at least 65,536$^3$, over a factor of $8$ larger than similar AMR simulations \citep{Collins2012,Li2015}.

\sloppy{
First, we drive turbulence until a quasi-steady-state is established after a few Eddy-turnover times. 
We then switch on self-gravity to let the cores collapse for approximately a free-fall time $t_{\rm ff}$. 
The collapse is approximately self-similar in this ideal MHD regime and the moving-mesh code continues to resolve the collapse down to smaller and smaller scales.  The collapse is stopped after a fraction of the free-fall time once ALMA-scale ($\sim$\,100\,AU) features are formed and resolved in the first cores that collapse (the simulations become prohibitively expensive after this point, as the simulation time-step is reduced as collapse proceeds). This happens at between approximately $0.2\,t_{\rm ff}$ (weak field) and $0.5\,t_{\rm ff}$ (strong field) in the simulations, at which point a few dozen cores have formed.  The simulations stop at the end of the isothermal collapse phase, but before the adiabatic collapse phase that would form the central protostar.  Nevertheless, we can still compare the simulations and the observations because---barring strong outflow feedback (see Section \ref{sec:disk_outflow})---the brief ($\sim$\,1000\,yr) period of adiabatic collapse that forms of the protostar should only affect the magnetic field morphology on <\,100\,AU scales, smaller than the $\sim$\,140\,AU resolution of our ALMA data.
}

The simulations are characterized by just three parameters: the sonic Mach number $\mathcal{M}_{s} = v_{\rm turb} / c_s$, the
virial parameter $\alpha_{\rm vir} = 5 v_{\rm turb}^2/(3G\rho_0L_0^2)$, and the Alfv\'enic Mach number of the mean magnetic field $\mathcal{M}_{{\rm A},\textrm{mean-field}}$. All the simulated clouds have virial parameter $\alpha_{\rm vir} = 1$, typical of GMC observations. In our Galaxy, giant molecular clouds are observed to be supersonic with $\mathcal{M}_{s}\sim 10$ (\citealt{Padoan2003}; note that some GMCs have values of $\mathcal{M}_{s} >10$), and with virial parameters $\alpha_{\rm vir}\sim 1$.  The magnetic field strength (characterized by $\mathcal{M}_{{\rm A},\textrm{mean-field}}$), on the other hand, is far more uncertain due to the limitations in observational techniques for measuring field strengths directly; therefore, we choose to investigate the effect of this single parameter.

The initial mass-to-flux ratio, parameterized in dimensionless form as $\mu_{\Phi,0}\equiv M_0/M_\Phi$, assesses the relative strength of the mean magnetic field and gravity. For our turbulent box, it can be approximated as $\mu_{\Phi,0}\sim \sqrt{5\pi/3}\,\mathcal{M}_{\rm A}$. Thus, our four simulations have $\mu_{\Phi,0}=80,8,2.7,0.8$. The strong-field case is estimated to be slightly sub-critical (i.e., the magnetic pressure is slightly greater than the gravitational energy), but we find that the turbulence can still drive a cloud core into collapse.  The three weak-field cases are super-critical, consistent with observations of magnetic field strengths in self-gravitating molecular clouds and the star-forming cores within them \citep{Myers1988, Crutcher2010,Crutcher2012}.

In the simulations, the masses of the star-forming cores are approximately $\left(8.8 \times c_s^2 \times R_{\textrm{core}}\right) / \,G \approx 4\,M_\odot$ for a thermal sound speed $c_s = 0.2$\,\kms and a core radius of $R_{\textrm{core}} =\,$\,10,000\,AU.  These values are comparable with the values found by \citet{Enoch2009b}, quoted above.\footnote{In the simulations, the density profile returns to the background level of density around $R_{\textrm{core}} =\,$\,10,000\,AU, beyond which radius any additional mass would be from background gas and dust.  As mentioned in Section \ref{sec:intro}, the mass estimates from \citet{Enoch2009b} are measured within 100,000\,AU due to the resolution of their observations; however, the majority of the mass to which their observations are sensitive is likely to be in a region $\ll 100,000$\,AU in extent.}

The simulations can be scaled to other physical parameters using the following relations (taking the mean molecular weight to be $2.3$\,amu):

{\footnotesize
\begin{align}
L_0 &= 5.2 \left(\frac{c_{\rm s}}{0.2~{\rm km}~{\rm s}^{-1}}\right)
\left(\frac{n_{\rm H}}{1000~{\rm cm}^{-3}}\right)^{-1/2}
\left(\frac{\mathcal{M}_{s}}{10}\right)~{\rm pc} \,\,, \\
B_0 &= 1.2,12,36,120 \left(\frac{c_{\rm s}}{0.2~{\rm km}~{\rm s}^{-1}}\right)
\left(\frac{n_{\rm H}}{1000~{\rm cm}^{-3}}\right)^{1/2}~{\rm \mu G} \,\,, \\
M &= 8000 \left(\frac{c_{\rm s}}{0.2~{\rm km}~{\rm s}^{-1}}\right)^3
\left(\frac{n_{\rm H}}{1000~{\rm cm}^{-3}}\right)^{-1/2} 
\left(\frac{\mathcal{M}_{\rm s}}{10}\right)^3 M_\odot \,\,.
\normalsize
\end{align}
}%

\smallskip
We have scaled our simulations to physical units using a sound speed of $c_{\rm s}=0.2~{\rm km}~{\rm s}^{-1}$ and hydrogen density $n_{\rm H} = 1000~{\rm cm}^{-3}$.  This density value is valid on the largest (5.2\,pc) scales of our simulations, and for a few $\times$ the largest eddy turnover time, which is 5.2\,pc\,/\,$(\mathcal{M}_{\rm s} c_{\rm s}) \approx$~2.5\,Myr.

The present work focuses on comparing the set of AREPO simulations with observations.  For a detailed analysis of the simulations, along with the theoretical properties of the collapsed protostellar cores (e.g., radial density and energy profiles, and the scaling of magnetic field with density), see \citet{Mocz2017}, who find that turbulent motions persist---and can still dominate the energy density---down to scales <\,10,000\,AU in the collapsing cores.

\vspace{1em}
\section{Results and analysis}
\label{sec:results}

In Figure~\ref{fig:3panel}(c), we show a map of the magnetic field orientation, inferred from polarized dust emission measured by ALMA, toward the young protostellar object Ser-emb~8.  This map has $\sim$\,7 times the resolution and more than an order of magnitude higher sensitivity than previous polarization observations of Ser-emb~8 by CARMA \citep{Hull2014, Hull2015b}.  The magnetic field revealed by the ALMA data does not have an hourglass morphology, and at the smallest scales exhibits perturbed structure that is not resolved in the lower-resolution observations by CARMA or the James Clerk Maxwell Telescope (JCMT); see Figure~\ref{fig:3panel}, panels (b) and (a), respectively.

\begin{figure*}
\begin{center}
\includegraphics[width=1\textwidth, clip, trim=0.5cm 5.5cm 3.5cm 0cm]{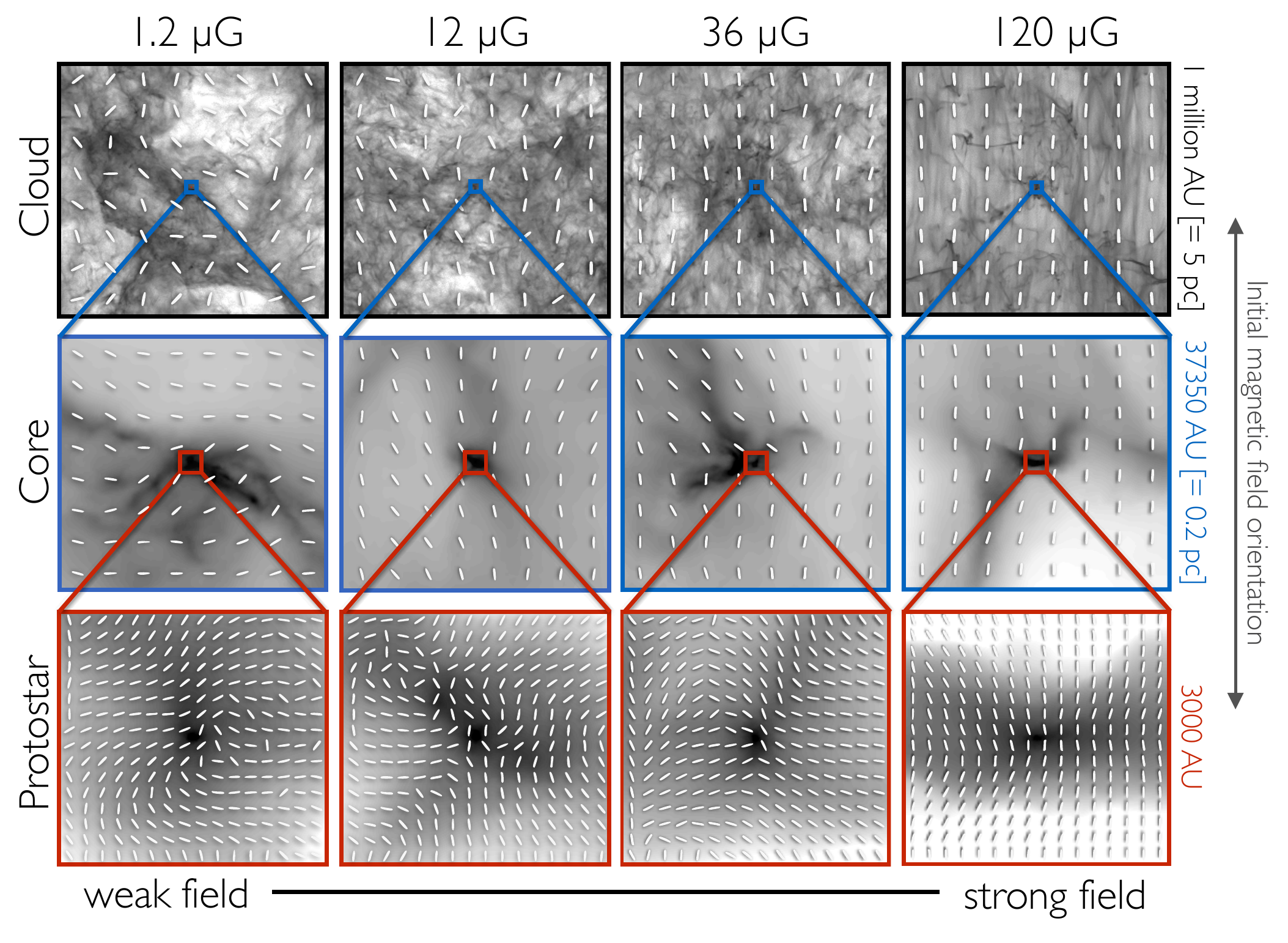}
\end{center}
\vspace{-1em}
\caption{\footnotesize Multi-scale projections of simulated data.  Shown are the column densities and the magnetic field orientation of protostellar cores that formed in our AREPO simulations. As indicated, the initial mean magnetic field is oriented in the vertical direction.  The initial magnetic field strength increases from the left to the right columns, corresponding to Alfv\'en Mach numbers $\mathcal{M}_{{\rm A},\textrm{mean-field}}=35,3.5,1.2,0.35$ (see Table \ref{table:sims}).  The top row shows the full-scale AREPO simulations, centered on the cores shown in the bottom row.  The full simulation boxes are 5.2\,pc in extent (cloud scale).  The middle row shows zoom-ins between JCMT and CARMA scales (core scale).  The bottom row shows zoom-ins of cores at ALMA scales (protostar scale). 
\textit{An interactive version of this figure is available in the online version of the ApJ Letter: by allowing the magnetic field orientations and zoom boxes to be toggled on and off, the interactive figure enables the reader to compare more easily the background grayscale column density maps with the foreground magnetic field orientations.}
}
\vspace{1em}
\label{fig:sim_proj}
\end{figure*}

While ALMA continuum dust polarization measurements reveal the dust intensity and magnetic field orientation in the plane of the sky (with caveats discussed in Section \ref{sec:grain_alignment}), they cannot be used to measure directly the strength of the magnetic field or the relative importance of other physical processes such as turbulence and gravity, which can be probed only with kinematic observations.  Simulations are therefore vital in order to achieve a deeper understanding of the underlying physics that leads to the observed density and magnetic field morphology at $\sim$\,100\,AU scales.  In Figure~\ref{fig:sim_proj} we show AREPO magnetohydrodynamic (MHD) simulations of star-forming cores in a turbulent medium that achieve the same resolution as the ALMA observations.

We find that changing the initial magnetic field strength in the AREPO simulations at the 5\,pc scale of the cloud dramatically alters the morphology of both the density and the magnetic field on spatial scales four orders of magnitude smaller (see Figure~\ref{fig:sim_proj}). 
In the strong-field case (i.e., the sub-Alfv\'enic case where magnetic energy dominates turbulent energy and $\mathcal{M}_{\rm A} < 1$), the field shapes the collapse, creating an obvious filamentary structure aligned perpendicular to the overall magnetic field orientation.  However, in the weak-field cases (i.e., the super- or trans-Alfv\'enic cases where turbulent energy dominates magnetic energy; $\mathcal{M}_{\rm A} > 1$) there are no clear, magnetically induced filaments; rather, the magnetic field is shaped by the dynamic properties of the gas, as expected for super-Alfv\'enic turbulence \citep{Burkhart2009}.  We note that in the three simulations with super- or trans-Alfv\'enic (weak-field) initial conditions, $\mathcal{M}_{\rm A} \approx 1$ at the $\sim$\,10$^4$\,AU scale of the collapsed cores.
Furthermore, the ALMA-scale simulations show significantly more fragmentation in the weak-field cases: this is consistent with the ALMA observations---which show several smaller companions near the main source---and confirms that fragmentation is more effectively suppressed in the strong-field case \citep[e.g.,][]{VazquezSemadeni2005,LewisBate2017}. 
These results argue that the relative importance of the magnetic field and turbulence at large ($\sim$\,5\,pc) scales is critical for determining the structure of a forming star all the way down to the $\sim$\,100\,AU spatial scales probed by our simulations and the ALMA observations.  

\vspace{1em}
\subsection{Histogram of Relative Orientation}

Recently, the \textit{Planck} satellite team assessed the role of the magnetic field in cloud dynamics by quantifying the relationship between the dust density structure and magnetic field orientation \citep{PlanckXXXII}.  The ``Histogram of Relative Orientation'' (HRO) method \citep{Soler2013} can be used to study how morphological density structures and the magnetic field are related. The method produces a histogram of relative orientations between the magnetic field and density gradient, and is based on computer vision techniques. 

The HRO is computed as follows. The relative alignment of the magnetic field $\textbf{B}$ and the local gradient of the density $n$ (or intensity $I$ in the observations) is characterized using the angle $\phi = \arctan{ \left(|\textbf{B} \times \nabla n|\right) / \left(\textbf{B} \cdot \nabla n\right) }$ between them. The histogram of $\phi$ for angles measured in 2 dimensions in the observations (or $\cos{\phi}$ for angles measured in 3 dimensions in the simulations) is known as the HRO. A flat histogram corresponds to random relative orientations. The histogram is constructed simply by calculating this angle for each pixel/voxel in the data/simulation and weighting the contribution from each element by a 1 or 0 depending on whether the magnitude of the density gradient is above the median value.  The \textit{Planck} team applied the method to high-, medium-, and low-density emission in the 15$\arcmin$ resolution \textit{Planck} data, and to mock simulations, where they found that, at the parsec scale of star-forming clouds, (1) at low densities the magnetic field is preferentially oriented parallel to density structures; (2) above a critical density the orientation changes from parallel to perpendicular; and (3) this change of relative orientation is the most significant if the magnetic field is strong.  

\begin{figure}
\begin{center}
\includegraphics[width=0.5\textwidth, clip, trim=0cm 3cm 0cm 0cm]{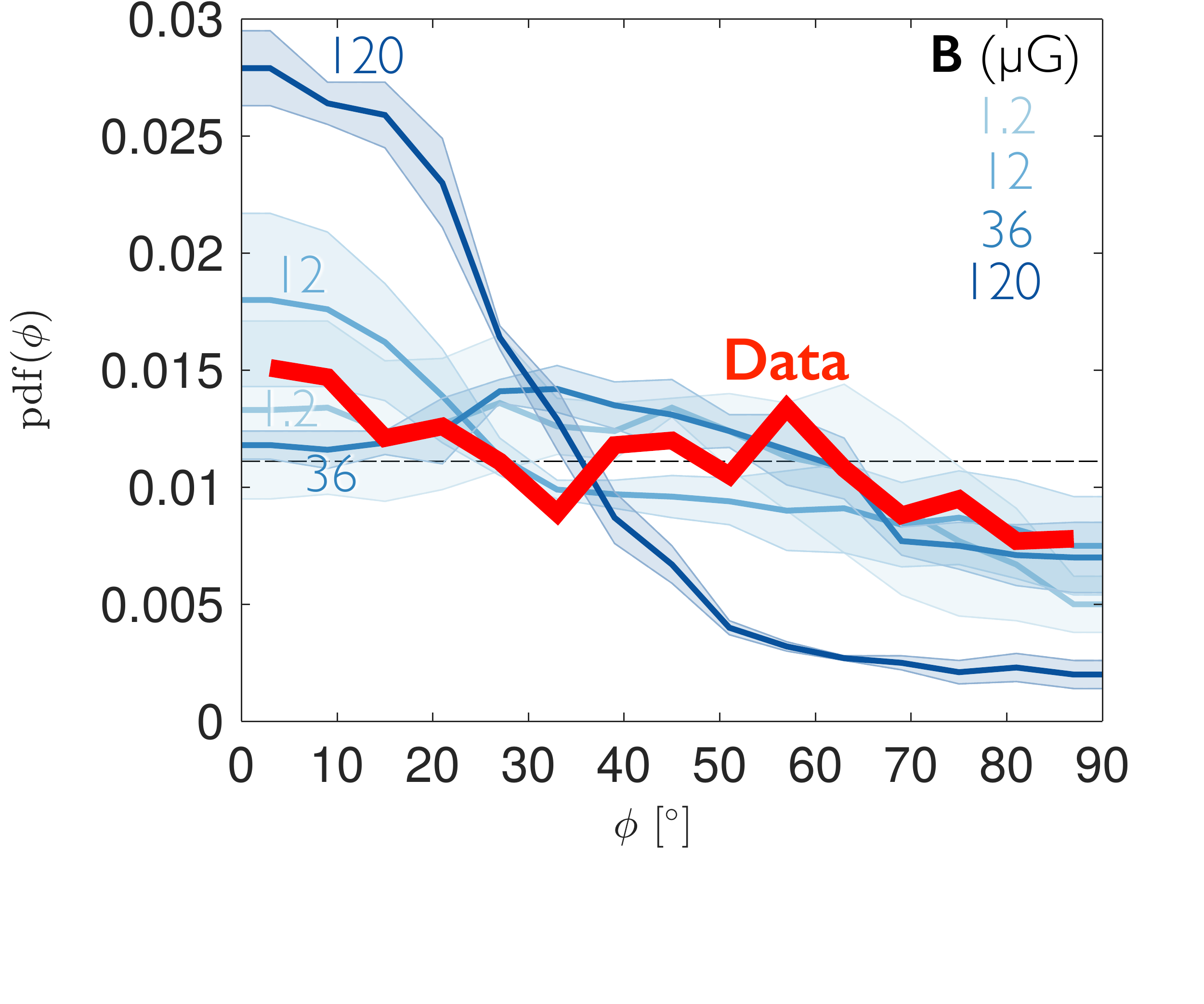}
\end{center}
\caption{\footnotesize Histogram of Relative Orientation (HRO): shown are the calculated HRO relations between the magnetic field orientation and the density gradients in the Ser-emb~8 ALMA data, as well as in the simulated data within a region of $3000~{\rm AU}$. Ser-emb~8 shows randomly distributed orientations, consistent with the weak-field simulations with $\mathcal{M}_{{\rm A},\textrm{mean-field}}>1$.  The shaded error regions reflect the variation in the multiple lines of sight used to calculate the HRO of each simulation.}
\label{fig:HRO}
\end{figure}

Here we use this technique to shed light on the role of the magnetic field at the $\sim$\,100\,AU scale of an individual star-forming core, calculating the HRO for both the ALMA observations and the AREPO simulations.  
For the observations, we assume the magnetic field has an orientation that is perpendicular to the measured polarization, and we take the gradient in the unpolarized dust emission (Stokes $I$), which traces column density structure. 
For the simulations, we calculate synthetic polarization in a $3000$\,AU region centered on each core assuming the optically thin limit.  We compute projected densities and polarization orientations on a uniform grid of cell length 12 AU, for various lines of sight.  The simulations themselves have an effective spatial resolution of $\sim$\,50\,AU on this spatial scale, which is comparable to the spatial resolution of the ALMA observations, and therefore no further convolution is applied in the mock observations (downsampling the mock projected column densities has no significant effect on the computed HRO).  The local polarization of a gas cell is $Q = \rho\cos(2\psi)\sin^2 i$, $U = \rho\sin(2\psi)\sin^2 i$, where $\rho$ is the gas density, $\psi$ is the local orientation of the magnetic field projected onto the plane of sky, and $i$ is the inclination angle of the local magnetic field relative to the line of sight. $Q$, $U$, and the column density are all integrated along the same line of sight. The HRO for each simulation is computed by averaging the results from multiple lines of sight, and is found to be largely independent of orientation: the shaded error regions in Figure \ref{fig:HRO} reflect the variation in the HROs calculated along different lines of sight.  

We explored computing the HRO for the ALMA data and simulated ALMA-like data as a function of different density cuts, as the \textit{Planck} team has done. However, we find that on ALMA scales, where the density of all of the gas is well above the critical density $\rho_{\rm crit}$ of collapse, neither the ALMA data nor the ALMA-like simulated cores with weak mean fields exhibit meaningful differences when the data are divided in a similar way. This is not unexpected, as we probe vastly smaller ($\sim$\,100\,AU) scales than $\lesssim 1$\,pc scales of the \textit{Planck} observations of nearby star-forming clouds. On the other hand, the strong-field simulation does exhibit a higher degree of alignment between the density gradient and magnetic field as density increases.

In Figure \ref{fig:HRO} we show the HRO for the ALMA observations and for all four simulations.  The HRO of the strongly magnetized simulation peaks at $\phi = 0\degree$, implying that the magnetic field is more systematically perpendicular than parallel to the filamentary structure traced by the dust.  This is in agreement with observations of filamentary star-forming structures that are thought to have a dynamically important magnetic field \citep{Pereyra2004, Alves2008, Goldsmith2008, Franco2010, Palmeirim2013, PlanckXXXII}
and with the observation that magnetic field orientation can be preserved across many orders of magnitude in spatial scale \citep{HBLi2009, Hull2014}.  However, we find that the HRO of the ALMA observations of Ser-emb~8 is flat, indicating a random field similar to the weakly magnetized simulations.  

The joint HRO analysis of our ALMA observations and AREPO simulations---and the differences in magnetic field morphology from $>$\,$80,000$\,AU cloud scales \citep{Matthews2009,Sugitani2010} to the $\sim$\,1000--100\,AU scales measured by CARMA and ALMA---lead us to conclude that Ser-emb~8 formed in a medium where turbulence at the cloud scale is super- or trans-Alfv\'enic and shapes the magnetic field at the smallest scales, causing the morphology of the field immediately surrounding the protostar to be disconnected from the history of the field on larger scales.

\section{Discussion}
\label{sec:disc}
\subsection{Consistency of the ALMA data with the weakly magnetized AREPO simulations}

In the sub-Alfv\'enic simulation (strongly magnetized; $\mathcal{M}_{{\rm A}}=0.35$), the magnetic field lines form an hourglass shape at small scales (see Figure \ref{fig:sim_proj}, bottom-right panel), which has been predicted by models \citep{Fiedler1993, Allen2003, Machida2005, Machida2006} and seen in observations of both low- and high-mass forming stars \citep{Girart2006, Girart2009, Rao2009, Tang2009, Qiu2013, Stephens2013, Qiu2014, HBLi2015}.  Neither the ALMA data nor the three weakly magnetized simulations show an obvious hourglass shape in the magnetic field, indicating that in a turbulent environment the hourglass morphology arises only in the strong-field case where the large-scale field is dynamically important. 

The formation of Ser-emb~8 deviates strongly from the idealized theoretical picture of a collapsing, quiescent, magnetically supported core with an hourglass-shaped field, suggesting that modern analytic models of star formation must take into account turbulent initial conditions. The random field morphology and the lack of an hourglass shape are evidence that the magnetic field in Ser-emb~8 is dynamically unimportant.

The dichotomy between sources with hourglass morphologies and those without implies a transition in the role of the magnetic field in the formation of stars in environments with sub-Alfv\'enic and super-Alfv\'enic turbulence.  Previous studies \citep{Hull2014} have hinted at a bimodality between the strong- and weak-field cases,\footnote{\citet{Hull2014} see different behavior in sources with low vs. high polarization fraction.  Polarization fraction may or may not be positively correlated with magnetic field strength; however, \citet{Hull2014} were unable to confirm this because the magnetic field strength cannot be measured directly from dust polarization observations, and because their CARMA observations had too few independent polarization detections to perform the HRO analysis we show in this work.} but lacked the resolution and the sensitivity to confirm that sources like Ser-emb 8 formed in environments dominated by turbulence.  Future high-resolution studies of magnetic fields in young, embedded protostellar sources will reveal whether stars tend to form in more weakly or strongly magnetized environments.

\subsection{Dust-grain alignment}
\label{sec:grain_alignment}

How well dust grains are aligned with the magnetic field depends on many factors including the grain population size and composition, the wavelength of the observations, and the density of the medium.  Furthermore, it has recently been predicted that polarization at (sub)millimeter wavelengths at scales of tens of AU within the high-density regions of protoplanetary disks can be caused by radiative grain alignment \citep{Tazaki2017} or by self-scattering of emission by large dust grains \citep{Kataoka2015}; there is now potential observational evidence for this self-scattering effect \citep{Kataoka2016b}, and other work showing that high-resolution CARMA and Karl G. Jansky Very Large Array (VLA) polarization observations \citep{Stephens2014,Cox2015,FernandezLopez2016} may be consistent with self-scattered dust emission \citep{Kataoka2016,Pohl2016,Yang2016a,Yang2016b, Yang2017}.  The ALMA polarization observations we present here are unlikely to be affected by these phenomena because the 140\,AU resolution data probe scales much larger than a typical protostellar disk.\footnote{Recent work has shown that disks in Class 0 protostellar sources tend to be small or unresolved: $R \sim 50$\,AU for L1527 (see \citealt{Ohashi2014}, and \citealt{Tobin2012} for earlier results); in other Class 0 sources, $R < 30$\,AU (see \citealt{Tobin2015b, SeguraCox2016})}

\begin{figure*}[hbt!]
\begin{center}
\includegraphics[width=0.50\textwidth, clip, trim=0cm 3cm 0cm 
0cm]{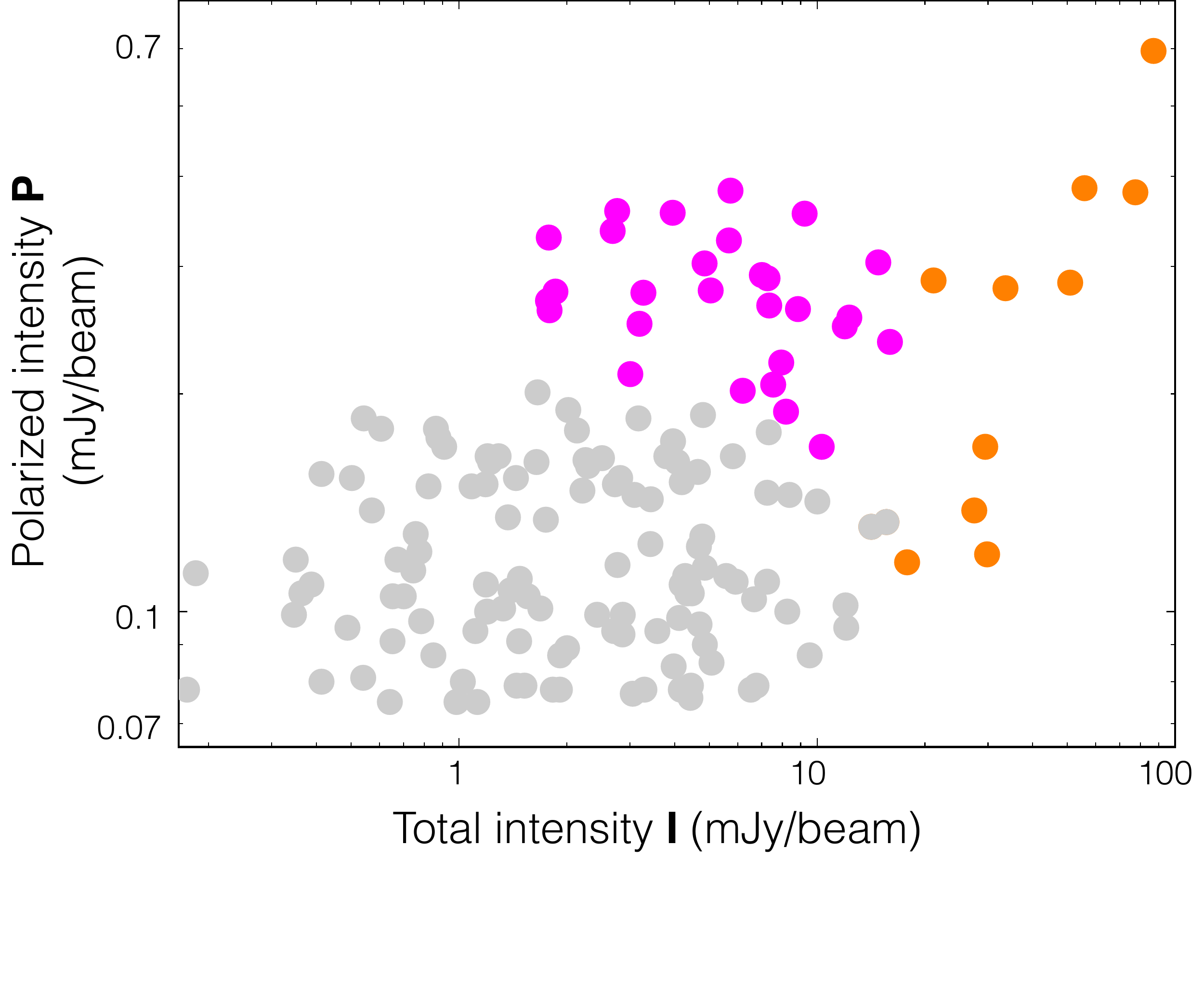}
\includegraphics[width=0.49\textwidth, clip, trim=0cm 2.6cm 0cm 0cm]{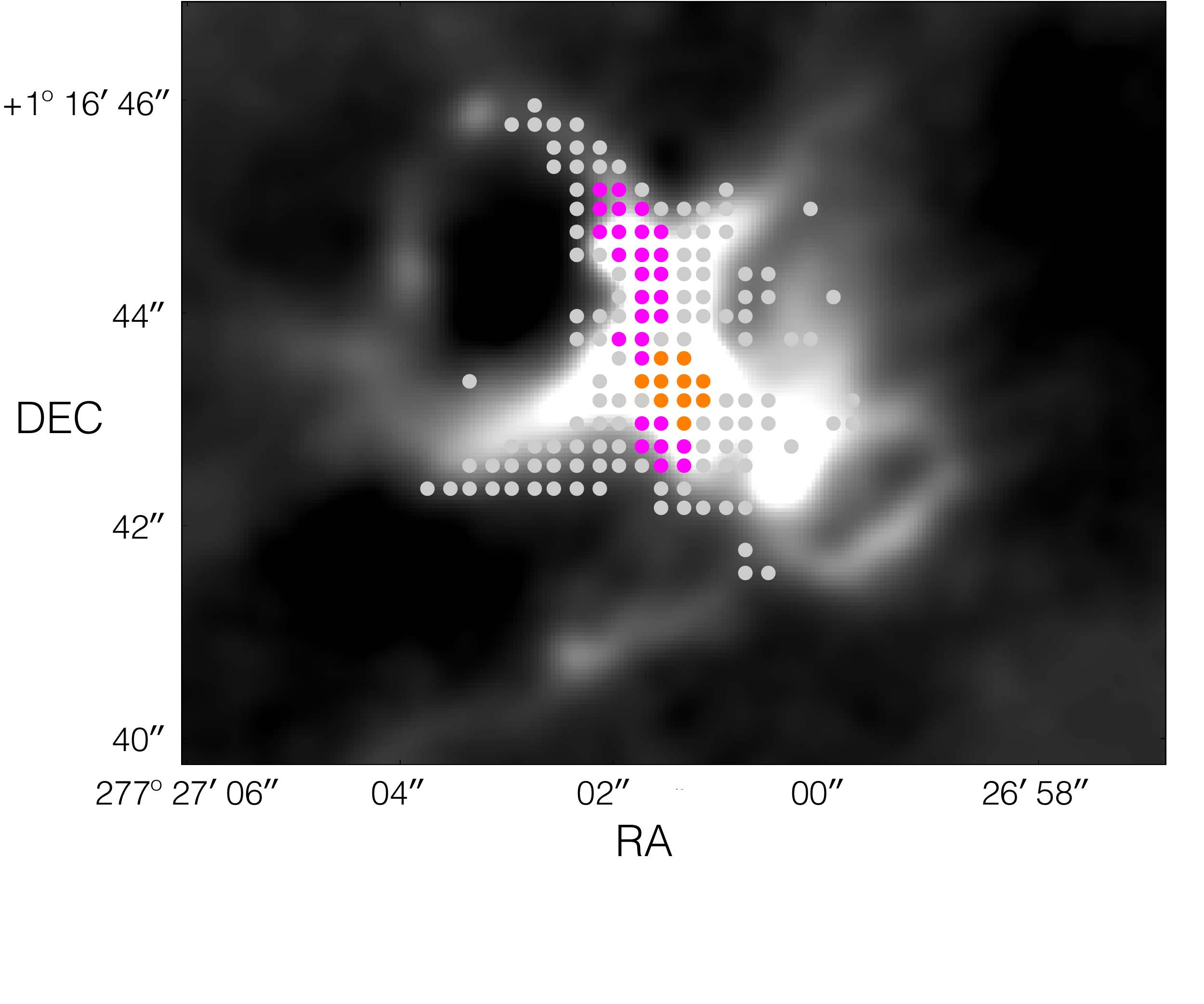}
\end{center}
\caption{\footnotesize Polarized intensity vs. total intensity in Ser-emb 8.  Values in the $P$ vs. $I$ plot on the left are plotted for pixels in the Ser-emb 8 ALMA map (shown on the right, and also in Figure \ref{fig:3panel}(c)) where there are significant detections of both $P$ and $I$ (see Table \ref{table:data}).  The gray, magenta, and orange points on the left-hand plot are associated with the points of the same color in the right-hand plot of Ser-emb~8.}
\label{fig:P_vs_I}
\vspace{0.1in}
\end{figure*}

\begin{table}[hbt!]
\normalsize
\begin{center}
\caption{\normalsize \vspace{0.1in} ALMA polarization data}
\begin{tabular}{cccccc}
\hline
\hline
$\alpha_{\textrm{J2000}}$ & $\delta_{\textrm{J2000}}$ & $\chi$ & $\delta\chi$ & $P$ & $I$\\
 & & (\degree) & (\degree) & \mjybmvert & \mjybmvert \\
\hline
277.45093 &   1.27777 &    46.8 &     8.9 &   0.081 &     --- \\ 
277.45009 &   1.27782 &    52.4 &     8.9 &   0.081 &     --- \\ 
277.44954 &   1.27793 &    70.6 &     8.6 &   0.083 &     --- \\ 
277.44998 &   1.27804 &    26.5 &     6.9 &   0.104 &     --- \\ 
277.45098 &   1.27810 &   171.4 &     8.1 &   0.088 &     --- \\ 
277.44993 &   1.27810 &    27.1 &     7.5 &   0.095 &     --- \\ 
277.45021 &   1.27821 &    51.1 &     7.8 &   0.091 &    0.652 \\ 
277.45015 &   1.27821 &    53.2 &     9.5 &   0.075 &    0.984 \\ 
277.45098 &   1.27827 &   146.5 &     8.7 &   0.082 &     --- \\ 
277.45093 &   1.27827 &   141.0 &     9.4 &   0.076 &     --- \\ 
277.45076 &   1.27827 &   156.7 &     8.8 &   0.082 &     --- \\ 
277.45021 &   1.27827 &    44.2 &     7.6 &   0.094 &    1.111 \\ 
277.45098 &   1.27832 &   142.7 &     8.0 &   0.090 &     --- \\ 
277.45093 &   1.27832 &   132.9 &     6.0 &   0.120 &     --- \\ 
277.45087 &   1.27832 &   139.0 &     8.6 &   0.083 &     --- \\ 
277.45082 &   1.27832 &   138.2 &     6.9 &   0.104 &     --- \\ 
277.45076 &   1.27832 &   144.1 &     8.0 &   0.090 &     --- \\ 
277.45098 &   1.27838 &   132.0 &     4.2 &   0.169 &     --- \\ 
277.45093 &   1.27838 &   126.9 &     4.1 &   0.174 &     --- \\ 
277.45087 &   1.27838 &   130.7 &     5.0 &   0.142 &     --- \\ 
277.45082 &   1.27838 &   130.6 &     5.6 &   0.128 &     --- \\ 
277.45076 &   1.27838 &   136.2 &     9.3 &   0.077 &     --- \\ 
277.45071 &   1.27838 &   148.9 &     7.3 &   0.098 &     --- \\ 
277.45065 &   1.27838 &   135.9 &     7.6 &   0.094 &     --- \\ 
277.45043 &   1.27838 &     7.4 &     6.0 &   0.118 &    0.673 \\ 
277.45037 &   1.27838 &    19.2 &     6.6 &   0.109 &    1.189 \\ 
277.45032 &   1.27838 &    32.0 &     6.4 &   0.111 &    1.478 \\ 
277.45026 &   1.27838 &    40.3 &     9.2 &   0.078 &    1.913 \\ 
277.45021 &   1.27838 &    46.7 &     9.2 &   0.078 &    3.294 \\ 
277.45104 &   1.27843 &   129.1 &     7.4 &   0.097 &    0.784 \\ 
277.45098 &   1.27843 &   132.8 &     4.4 &   0.161 &    1.221 \\ 
277.45093 &   1.27843 &   131.4 &     5.3 &   0.135 &    1.370 \\ 
277.45087 &   1.27843 &   136.5 &     4.8 &   0.149 &    1.084 \\ 
277.45082 &   1.27843 &   136.7 &     4.8 &   0.149 &    0.822 \\ 
277.45076 &   1.27843 &   135.9 &     5.9 &   0.121 &    0.775 \\ 
277.45071 &   1.27843 &   141.3 &     4.0 &   0.179 &    0.606 \\ 
277.45065 &   1.27843 &   135.9 &     4.6 &   0.155 &    0.413 \\ 
277.45059 &   1.27843 &   136.8 &     9.0 &   0.080 &    0.413 \\ 
277.45043 &   1.27843 &   176.1 &     4.5 &   0.161 &    1.646 \\ 
277.45037 &   1.27843 &     1.4 &     4.7 &   0.153 &    2.819 \\
277.45093 &   1.27849 &   134.3 &     6.2 &   0.116 &    2.771 \\ 
277.45087 &   1.27849 &   133.5 &     4.8 &   0.150 &    2.725 \\ 
277.45082 &   1.27849 &   129.5 &     4.4 &   0.162 &    2.255 \\ 
... &   ... &   ... &     ... &   ... &    ... \\ \\
\hline
\vspace{-1.5em}
\end{tabular}
\label{table:data}
\end{center}
\footnotesize
\textbf{Note.} $\chi$ is the orientation of the magnetic field, measured counterclockwise from north.  $\delta\chi$ is the uncertainty in the magnetic field orientation.  $P$ is the polarized intensity.  $I$ is the total intensity, reported where $I > 3\sigma_I$.  Due to differences in dynamic range between the images of Stokes $I$ and polarized intensity, there are cases where $P$ is detectable but $I$ is not.  \textit{The full, machine-readable table is available in the online version of the ApJ Letter.}
\end{table}

We turn now to the question of whether the grains deep inside the protostellar envelope are aligned by the magnetic field.  Because interstellar photons cannot penetrate into the coldest, densest regions we are probing with ALMA, a bright protostar must provide the radiation flux from within the source in order to align the dust grains. We know that there is a protostar at the center of the system, most clearly because we see a bipolar outflow being launched by the source (see \citealt{Hull2014}, Figure 26a); the emission from this source will serve to align the dust grains deep within Ser-emb 8.  Furthermore, a benefit of observing with an interferometer is that the larger-scale emission is filtered out, so the CARMA and ALMA data are not sensitive to the larger-scale JCMT data. Therefore, the polarized emission that we see in the ALMA map is from aligned dust grains at spatial scales between approximately 140 AU and a few thousand AU (i.e., the largest-scale structures recovered by the ALMA data).

One way to test the potential effects of grain alignment in our map of Ser-emb~8 is to look at the behavior of the polarized intensity $P$ vs. the total intensity $I$ \citep{Arce1998}.  A number of authors have performed this type of test, although they tend to plot the polarization fraction $P/I$ vs. $I$ \citep{Cho2005, Bethell2007, Pelkonen2009, Alves2014, Andersson2015}; we choose to plot $P$ vs. $I$ in order to avoid plotting a function of $I$ against itself.  We do so in Figure \ref{fig:P_vs_I} (left); see also the corresponding points on the map of Ser-emb~8 in the right-hand plot.  

Of the 10 orange points, which correspond to the locations nearest to the total intensity $I$ peak of the source, most of them also have among the highest polarized intensities $P$ detected toward the source.  This indicates that there is additional polarized radiation originating from the very center of the source, suggesting that grains in the very interior of Ser-emb 8 are aligned by the radiation from the embedded protostar, and that polarization may still be a faithful tracer of magnetic field morphology even down to $\sim$\,100\,AU scales.  However, the trends in Figure \ref{fig:P_vs_I} are weak, and could be due to geometrical effects.  For example, the magenta points have high $P$ but lower $I$ and are clustered in discrete areas, suggesting that these may be regions where the magnetic field happens to be oriented close to the plane of the sky, which would explain the increase in polarized emission.  The remainder of the points (in gray) comprise a scatter plot, where $P$ is approximately constant with $I$ across the majority of the source.  This is consistent with super- or trans-Alfv\'enic turbulence, where a randomly oriented magnetic field is integrated along the line of sight.  This is true even under the assumption of perfect grain alignment \citep{FalcetaGoncalves2008}.

We should note that even if the dust grains deep within the protostellar core were not magnetically aligned, we would still arrive at the conclusion that Ser-emb~8 formed in a weakly magnetized environment.  This is because the polarization signal we detect in our ALMA data would then be coming from only the aligned grains in the lower-density region of the core further away from the central protostar.  If that were the case, then the ALMA observations would show that the magnetic field morphology even in the low-density material---where the field is less likely to be affected by small-scale dynamics---is inconsistent with strong-field star-formation models.

\subsection{Impact of disks and outflows}
\label{sec:disk_outflow}

While the AREPO simulations we present here do not form outflows or Keplerian disks, it is possible that in some sources the magnetic field morphology could be affected by disk rotation and outflow feedback.  
Regarding outflows, in Ser-emb 8 the outflow is known to be well collimated \citep{Hull2014}, and thus the disturbance of the field by the outflow should be minimal because the solid angle subtended by the outflow is much smaller than the core and envelope regions traced by the JCMT, CARMA, and ALMA maps.  Furthermore, past studies have successfully modeled the hourglass structure of NGC~1333-IRAS~4A \citep{Goncalves2008, Frau2011} and IRAS~16293A \citep{Rao2009}---both of which have very powerful bipolar outflows---and have obtained excellent fits to the observations without explicitly including the influence of disks or outflows.   
Nevertheless, we acknowledge that the impact of disks and outflows may range beyond the physical extent of those structures due to the propagation of disturbances in the field in the form of Alfv\'en waves from outflow propagation \citep{DeColle2005,Frank2014} and disk rotation \citep{Matsumoto2004, Kataoka2012}.  

The fact that outflows and magnetic fields are randomly aligned in both observations \citep{Hull2013} and comparable synthetic observations of simulations \citep{JLee2017} is intriguing because it suggests that both outflows and disks are largely unaffected by---and thus behave independently of---the magnetic fields we examine here.  Future ALMA polarization observations at even smaller scales (e.g., well-resolved polarimetric observations of Class 0 disks improving upon studies by, e.g., \citealt{Rao2014,Stephens2014,SeguraCox2015}) will reveal whether the interplay among outflows, disks, and magnetic fields changes fundamentally at the small scales where disks are formed and outflows are launched.

\section{Conclusions}
\label{sec:conc}

We report the first ALMA observations of polarized dust emission from the protostellar source Ser-emb 8 at a linear resolution of 140\,AU. Contrary to the expectation of models of strongly magnetized star formation, the magnetic field in Ser-emb 8 neither exhibits an hourglass morphology nor is consistent with the large-scale magnetic field.

In order to study the behavior of the magnetic field's orientation, we ran four magnetohydrodynamic gravo-turbulence simulations run with the AREPO code that have initial conditions ranging from super-Alfv\'enic (weakly magnetized) to sub-Alfv\'enic (strongly magnetized). We resolve the collapse of a molecular cloud from the parsec scale of star-forming clouds down to the $\sim$\,100\,AU scale probed by ALMA, and find that:

\begin{enumerate}

\item Only in the very strongly magnetized (sub-Alfv\'enic) simulation ($\sim$\,120\,$\mu$G on the 5.2\,pc cloud scale) do we see both the preservation of the field direction from cloud to disk scales and an hourglass-shaped field at <\,1000\,AU scales.

\item  When the cloud-scale magnetic field is weak (super- or trans-Alfv\'enic), turbulence shapes the field on small scales, divorcing it from the mean large-scale field.

\item An HRO analysis of the AREPO simulations reveals that only the strongly magnetized simulation has density gradients and magnetic fields that are aligned on $\sim$\,100\,AU scales.  When the cloud is weakly magnetized, the density and magnetic field orientation are randomly aligned.  The latter simulations are consistent with the ALMA data.

\end{enumerate}

Based on the comparison with simulations, we conclude that the turbulence in the natal cloud of Ser-emb 8 is super- or trans-Alfv\'enic, as it exhibits a random alignment of field and intensity, in contrast with the sub-Alfv\'enic simulation, where the magnetic field plays a role in dictating the formation of the source structure.  In this weak-field scenario, turbulent energy dominates magnetic energy, and thus turbulence is able to shape the material that forms the protostar across many orders of magnitude in spatial scale.  With the powerful combination of high sensitivity, resolution, and spatial dynamic range; and with the statistical diagnostics provided by ALMA observations and AREPO simulations, it is now possible for us to obtain a multi-scale view of the role of the magnetic field in different star forming environments.

\acknowledgements

\sloppy{
The authors thank the anonymous referees, whose comments improved the manuscript substantially.
C.L.H.H. acknowledges the outstanding calibration and imaging work performed at the North American ALMA Science Center by Crystal Brogan, Jennifer Donovan Meyer, and Mark Lacy.
The authors acknowledge Juan Soler for the helpful discussion about the Histogram of Relative Orientation.
P.M. is supported in part by a NASA Earth and Space Science Fellowship. B.B. is a NASA Einstein Fellow. P.M. and B.B. thank Alex Lazarian, Chris McKee and Richard Klein for valuable discussions.
J.M.G. acknowledges support from MICINN AYA2014-57369-C3-P and the MECD PRX15/00435 grants (Spain).
Support for CARMA construction was derived from the states of California, Illinois, and Maryland, the James S. McDonnell Foundation, the Gordon and Betty Moore Foundation, the Kenneth T. and Eileen L. Norris Foundation, the University of Chicago, the Associates of the California Institute of Technology, and the National Science Foundation.
This Letter makes use of the following ALMA data: ADS/JAO.ALMA\#2013.1.00726.S. 
Z.-Y.L. is supported in part by AST 1313083 and NASA NNX14AB38G.
ALMA is a partnership of ESO (representing its member states), NSF (USA) and NINS (Japan), together with NRC (Canada), NSC and ASIAA (Taiwan), and KASI (Republic of Korea), in cooperation with the Republic of Chile. The Joint ALMA Observatory is operated by ESO, AUI/NRAO and NAOJ. 
The National Radio Astronomy Observatory is a facility of the National Science Foundation operated under cooperative agreement by Associated Universities, Inc.
The computations in this Letter were run on the Odyssey cluster supported by the FAS Division of Science, Research Computing Group at Harvard University.
This research made use of APLpy, an open-source plotting package for Python hosted at \url{http://aplpy.github.com}.
This work also made use of the data-visualization software Glue (\url{http://glueviz.org}, \url{https://github.com/glue-viz}); the authors acknowledge Penny Qian for rapid-response Glue development work and for her assistance in using the software.
}


\end{document}